\documentclass[10pt]{article}
\usepackage[utf8]{inputenc}
\usepackage[T1]{fontenc}
\usepackage{graphicx}
\usepackage{multirow}
\usepackage{amsmath}
\usepackage{float}
\usepackage{adjustbox}
\usepackage{authblk}
\usepackage{booktabs}
\usepackage{xcolor}
\usepackage{url}
\usepackage[margin=1in]{geometry}
\usepackage{subcaption}
\usepackage[superscript,nomove]{cite}
\usepackage{hyperref}
\hypersetup{
  colorlinks=true,
  linkcolor=blue,
  citecolor=blue,
  urlcolor=blue
}
\title{ERDES: A Benchmark Video Dataset for Retinal Detachment and Macular Status Classification in Ocular Ultrasound}

\author[1,*]{Yasemin Ozkut}
\author[1]{Pouyan Navard}
\author[2]{Srikar Adhikari, MD}
\author[2]{Elaine Situ-LaCasse, MD}
\author[2]{Josie Acu\~{n}a, MD}
\author[2]{Adrienne A Yarnish, MD}
\author[1]{Alper Yilmaz}

\affil[1]{PCVLab, The Ohio State University, USA}
\affil[2]{Department of Emergency Medicine, University of Arizona, USA}
\affil[*]{Corresponding author email: ozkut.1@osu.edu}

\date{}

\begin{document}
\sloppy
\maketitle
\begin{abstract}

Retinal detachment (RD) is a vision-threatening condition that requires prompt intervention to preserve sight. A critical factor in treatment urgency and visual prognosis is macular involvement---whether the macula is intact or detached. Point-of-care ultrasound (POCUS) is a fast, non-invasive and cost-effective imaging tool commonly used to detect RD in various clinical settings. However, its diagnostic utility is limited by the need for expert interpretation, especially in resource-limited environments. Deep learning has the potential to automate RD detection on ultrasound, but there are no clinically available models, and prior research has not addressed macular status---an essential distinction for surgical prioritization. Additionally, no public dataset currently supports macular-based RD classification using ultrasound video. We introduce \textbf{E}ye \textbf{R}etinal \textbf{DE}tachment ultra\textbf{S}ound (\textbf{ERDES}), the first open-access dataset of ocular ultrasound clips labeled for (i) presence of RD and (ii) macula-detached vs. macula-intact status. ERDES enables machine learning development for RD detection. We also provide baseline benchmarks by training 40 models across eight architectures, including 3D convolutional networks and transformer-based models.

\end{abstract}

\section*{Background \& Summary}

Retinal detachment (RD) is a serious ocular condition in which the neurosensory retina separates from the underlying retinal pigment epithelium (RPE), potentially leading to permanent vision loss if not treated immediately~\cite{Steel2014, GhaziGreen}. It affects approximately 1 in 10,000 people annually, with a lifetime risk of close to 1 in 300, and is more common in individuals with risk factors such as increasing age, prior cataract surgery, myopia, or ocular trauma~\cite{GarianoKim}. Common symptoms include flashes of light (photopsia), floaters, blurred vision, and loss of peripheral visual field~\cite{GarianoKim}. Given its potential for irreversible damage, RD is considered an ophthalmic emergency~\cite{GhaziGreen}. However, vision loss is often preventable with timely surgical intervention, especially when the macula---the central region of the retina responsible for high-acuity vision---remains attached~\cite{GarianoKim}. A closely related condition, posterior vitreous detachment (PVD), is a common age-related process in which the vitreous gel separates from the retina due to vitreous liquefaction and weakening of vitreoretinal adhesions, often causing symptoms that closely mimic the presenting symptoms of retinal detachment. However, unlike RD, uncomplicated PVD generally has a benign course, although it increases the risk of retinal tears that can progress to retinal detachment. As a result, the clinical presentation does not reliably distinguish a benign PVD from a vision-threatening RD~\cite{Johnson2010, AhmedTripathy2023, Hollands2009}. In developing this dataset, we focused on two primary diagnostic categories: eyes without retinal detachment (Non-RD), which includes both normal eyes and posterior vitreous detachment (PVD), and eyes with retinal detachment (RD), which are further subcategorized by macular status (macula-detached vs. macula-intact)---the latter being a key determinant of both prognosis and urgency of intervention.

The condition of the macula, particularly the fovea, is the strongest predictor of final visual acuity in cases of retinal detachment~\cite{Wykoff}. Retinal detachments are typically classified according to macular involvement: intact macula, when the central retina remains attached to the RPE, and detached macula, when this region has separated. Patients with macula-intact detachments generally retain better postoperative visual outcomes with urgent intervention, while those with macula-detached detachments often experience poorer visual recovery even after successful reattachment~\cite{Steel2014, Subramony2021}. As a result, early detection of macula-intact RD is essential to preserve central vision.

Macula-intact retinal detachments can progress rapidly, with subretinal fluid reaching the fovea and causing irreversible vision loss within hours~\cite{Steel2014, Mahmoudi2016, Subramony2021, Wykoff}. Clinical guidelines recommend surgical intervention on the same day, ideally within 24 hours, to prevent macular involvement~\cite{GhaziGreen, Grabowska2021}. In contrast, once the macula has detached, the prognosis worsens regardless of the surgical timing~\cite{Steel2014}. Consequently, macula-detached cases are typically scheduled within one to two weeks, with the aim of salvaging any remaining visual function~\cite{Mahmoudi2016, Subramony2021, Grabowska2021, GhaziGreen}. This management approach underscores the status of the macula as the key determinant of surgical evaluation in RD.

Ultrasound provides a non-invasive method to visualize intraocular structures and has been shown to be very accurate for the detection of RD~\cite{WangRizzuti}. This work focuses on point-of-care ultrasound (POCUS), which offers a rapid, accessible, and cost-effective imaging solution in diverse clinical settings~\cite{Subramony2021, WangRizzuti, Zieleskiewicz2021}. On ultrasound, RD typically appears as a bright, undulating, free-floating membrane within the vitreous tethered to the optic nerve~\cite{WangRizzuti}. In contrast, PVD typically appears as a thinner, more mobile, less echogenic membrane that is not firmly tethered to the optic disc, unlike the thicker, brighter, undulating membrane of an RD that remains attached at the optic nerve head~\cite{Propst2020}.

Despite its diagnostic value, ultrasound interpretation presents challenges. Accurate interpretation of ocular ultrasound clips and differentiation between RD and other ocular pathologies require expert recognition of non-RD sonographic anatomy and pathology. This skill level is currently not available universally among healthcare providers, which is a major limitation in the routine use of POCUS to detect RD. These limitations motivate the need for curated ultrasound video data that support consistent clip-wise labeling and machine learning–based analysis. 

Recent advances in deep learning have enabled new possibilities in automated medical image interpretation, including RD detection~\cite{Christ2024, Caki2025, Yadav2022}. However, most existing approaches focus solely on identifying RD and do not account for macular involvement, a clinically relevant distinction. Although one model has incorporated macular status~\cite{Li2020}, it relied on ultra-wide field (UWF) fundus photographs rather than ultrasound video. Although deep learning has been applied to ocular ultrasound in limited contexts, currently no openly accessible dataset supports macular-based RD classification from ultrasound clips (Table~\ref{tab:dataset_comparison}). 

\begin{table}[ht]
\centering
\caption{Comparison of existing datasets related to retinal detachment detection and classification.}
\label{tab:dataset_comparison}
\resizebox{\textwidth}{!}{%
\begin{tabular}{lcccccc}
\hline
\textbf{Study} & \textbf{Modality} & \textbf{Data Type} & \textbf{Size} & \textbf{Task} & \textbf{Macular Status} & \textbf{Access} \\
\hline
Li et al. (2020)\cite{Li2020}             & UWF Fundus                & Images           & 10,451 & RD detection, macula-on/off discerning & Yes & Upon Request \\
Chen et al. (2021)\cite{Chen2021}         & B-scan Ultrasound         & Images           & 4,521  & RD, VD, VH screening + segmentation & No  & Not Stated \\
Ye et al. (2024)\cite{Ye2024}             & B-scan Ultrasound         & Images           & 6,054  & RD, VH, IOT, PSS screening & No  & Upon Request \\
Caki et al. (2025)\cite{Caki2025}         & B-scan Ultrasound         & Images           & 279    & RD, PVD segmentation & No  & Not Stated \\
Wang et al. (2025)\cite{Wang2025}         & B-scan Ultrasound         & Images + Reports & 4,858 & Report generation (15 diseases) & No  & Not Public \\
Toyama et al. (2026)\cite{Toyama2026}     & B-scan Ultrasound         & Images           & 3,230  & RD, VH detection & No  & Upon Request \\
\hline
\textbf{ERDES (Ours)}                       & \textbf{Linear-Probe Ultrasound (POCUS)} & \textbf{Video clips} & \textbf{5,381} & \textbf{RD detection, macular status} & \textbf{Yes} & \textbf{Open (CC-BY 4.0)} \\
\hline
\end{tabular}%
}
\end{table}

To our knowledge, ERDES represents the first openly accessible ocular ultrasound dataset designed for RD triage using linear-probe POCUS video clips, with labels supporting both RD detection and macular status classification. To address this gap, the key contributions of this work are as follows:
\begin{enumerate}
  \item \textbf{ERDES Dataset:} We release \textbf{E}ye \textbf{R}etinal \textbf{DE}tachment ultra\textbf{S}ound, ERDES~\cite{ozkuterdes}, the first open-access collection of ocular ultrasound clips labeled for (i) absence vs.\ presence of retinal detachment (RD) and (ii) macula-detached vs.\ macula-intact status.

  \item \textbf{Comprehensive model benchmarking:} We trained and evaluated 40 models in eight spatiotemporal architectures for five binary tasks, focusing on the two primary tasks: Non-RD vs.\ RD and Macula-Detached vs.\ Macula-Intact. The architectures included 3D ResNet~\cite{3dresnet}, 3D U-Net~\cite{cicekunet3d}, V-Net~\cite{milletari2016v}, UNet++~\cite{zhou2018unet++}, SENet154~\cite{hu2018senet}, Swin-UNETR~\cite{hatamizadeh2021swin}, UNETR~\cite{hatamizadeh2022unetr}, and ViT-based~\cite{dosovitskiy2020image} classifiers. We report metrics including accuracy, precision, sensitivity, specificity, and F1-score.

  \item \textbf{Two-stage diagnostic pipeline:} Based on the benchmarking results, we designed a cascaded classification pipeline: the first stage model distinguishes RD from Non-RD clips, and the second stage model classifies RD-positive clips as Macula-Detached or Macula-Intact.

  \item \textbf{Open-source release:} To accelerate research in automated interpretation of ophthalmic ultrasound, we publicly release the ERDES dataset, labels, training scripts, and pre-trained models under a permissive license.
\end{enumerate}

\section*{Methods}
This section documents the creation of the ERDES ocular ultrasound collection, covering participant information, ultrasound acquisition, clip preparation, class labeling, and preprocessing. 

\subsection*{Ethics}
This study was approved by the Institutional Review Board (IRB) of the University of Arizona (IRB Parent Protocol ID: 2003423113). Ultrasound data were obtained exclusively from an existing archival database at our institution, with all datasets de-identified and no protected health information (PHI) accessed. As no live patient interaction occurred and the study relied solely on retrospective anonymized data, the IRB granted a waiver of consent.

\subsection*{Participants}

Although no identifiable demographic information was collected at the individual level, the ERDES dataset includes a racially diverse adult population ($\geq$ 18 years) of both sexes, including Caucasians, African Americans, Hispanics, American Indians, and Asian Americans.

\subsection*{Data Acquisition}
The ERDES dataset spans 2010–2022 and includes an average of approximately 250 cases per year. Ocular ultrasound clips were acquired using a variety of ultrasound devices: Mindray/Zonare (~45\%), Philips (~30\%), GE (~20\%), and Sonosite (~5\%). High-frequency linear broadband array transducers, ranging from 5 MHz to 12 MHz, were used to perform ocular ultrasound examinations. The scans were performed by POCUS trained physicians at the University of Arizona, in the Emergency Department setting---as part of routine clinical care for patients with visual symptoms. A standardized scanning protocol was followed for image acquisition. The patients were positioned supine to semi-upright and asked to look straight ahead with their eyes closed in a neutral position. A water-soluble ultrasound gel was applied over the closed eyelid, and the sonographers ensured that no pressure was applied to the globe. A comprehensive three-dimensional evaluation of the eye was performed by capturing images in two orthogonal planes and instructing patients to look in various directions (up, down, left, right) to optimize visualization. Both eyes were scanned identically to allow for direct comparison.

\subsection*{Clip Preparation}
Ocular ultrasound examinations were identified and extracted from Qpath, an image archiving and workflow management database. Qpath was queried specifically for ocular ultrasound studies, and the corresponding image clips were exported directly from the platform in MP4 format. During this export process, the platform automatically crops out all Protected Health Information (PHI), ensuring that the resulting video clips are fully deidentified. This automated deidentification and export process adheres to all HIPAA standards for data privacy and security.

\subsection*{Class Labelling}

Labels were generated using an adjudication-based workflow. Three clinical experts independently reviewed the ocular ultrasound clips and classified each clip into one of the following diagnostic categories: Non-RD or RD. Non-RD cases are subdivided into Normal and Posterior Vitreous Detachment (PVD), while RD cases are subdivided into Macula-Intact and Macula-Detached.

The labeling criteria are as follows: A normal eye on ultrasound shows the retina as a smooth, thin echogenic line closely opposed to the posterior globe/choroid, without any mobile membranes. PVD is seen as a thin, mobile echogenic membrane within the vitreous that is not tethered to the optic nerve and demonstrates marked mobility with ocular movements. The ultrasound findings of RD include the presence of a bright, uniformly thick, echogenic membrane tethered to the optic nerve (but separated from the choroid) that does not cross the optic nerve and is less mobile with ocular movements. In the presence of RD, macular involvement is determined by whether the detachment extends to the macular region. In ocular ultrasound, the optic nerve serves as a landmark to determine the location of the macula, and the macula lies temporal to the optic disc. Macula-intact retinal detachment refers to the cases where the detached retinal membrane on the temporal side of the optic nerve does not extend into the macular region. Retinal detachments confined to the nasal side of the optic nerve are considered macula-intact. Macula-detached retinal detachment refers to cases where the detached retinal membrane extends into the macular region.

Distinguishing Normal vs. PVD on ultrasound is more challenging than distinguishing Normal vs. RD because PVD findings are often subtle: the detached vitreous cortex may have low echogenicity, fine strands, or only faint mobile echoes that can be easily overlooked or misinterpreted, especially in suboptimal gain or noisy images. In contrast, RD usually produces a conspicuous, high-contrast, V- or Y-shaped, highly echogenic membrane that stands out clearly against the anechoic vitreous cavity, making Normal vs. RD discrimination more straightforward than Normal vs. PVD. The overlap in sonographic appearance, combined with operator dependence and variability in image acquisition, makes reliable differentiation between normal vitreous and PVD significantly more difficult than identifying RD. In addition, a thin, freely mobile echogenic PVD membrane may be mistaken for a thicker, more tethered RD unless attention is paid to key distinguishing features---for example, PVD often crosses the midline and is not anchored to the optic disc, such that small errors in probe angle, depth, or gain can blur these distinctions and lead to misclassification~\cite{Baker2018}.

Labeling was performed at the clip level, which means that each ultrasound video clip was assigned a single overall label rather than annotating individual frames. This approach was selected to reflect how ocular ultrasound is interpreted clinically and to account for normal temporal variation within a clip (e.g., eye motion and probe movement). Frame-level annotations were not performed.

Following the initial classification, a fourth expert reviewed all labeled clips to verify the accuracy and consistency of the classifications, serving as a quality control step to ensure label reliability. Inter-rater agreement statistics (e.g., Cohen’s/Fleiss’ kappa) were not calculated because labeling was finalized through adjudication rather than independent overlapping annotation. Reviewer-specific identifiers and consensus/disagreement metadata were not captured as part of the stored label metadata; the dataset contains the final validated label for each clip.

\subsection*{Pre-processing}

\begin{figure}[H]
  \centering
  \includegraphics[width=0.8\textwidth]{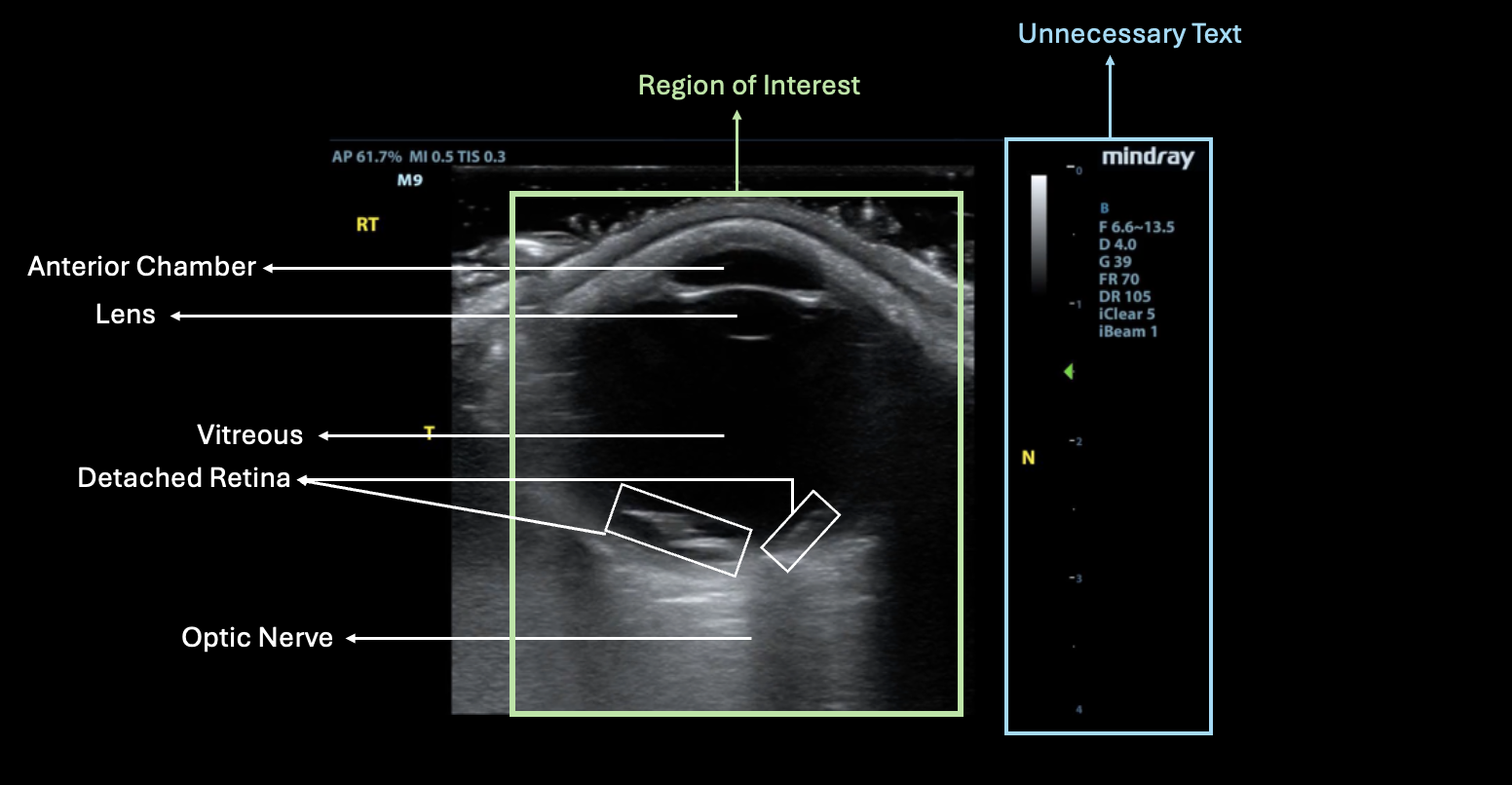}
  \caption{Illustration of ocular globe, peripheral soft tissue and device-generated text overlays.}
  \label{fig:eye_anatomy}
\end{figure}

\begin{figure}[ht]
  \centering
  \begin{subfigure}[b]{0.45\textwidth}
    \includegraphics[width=\textwidth]{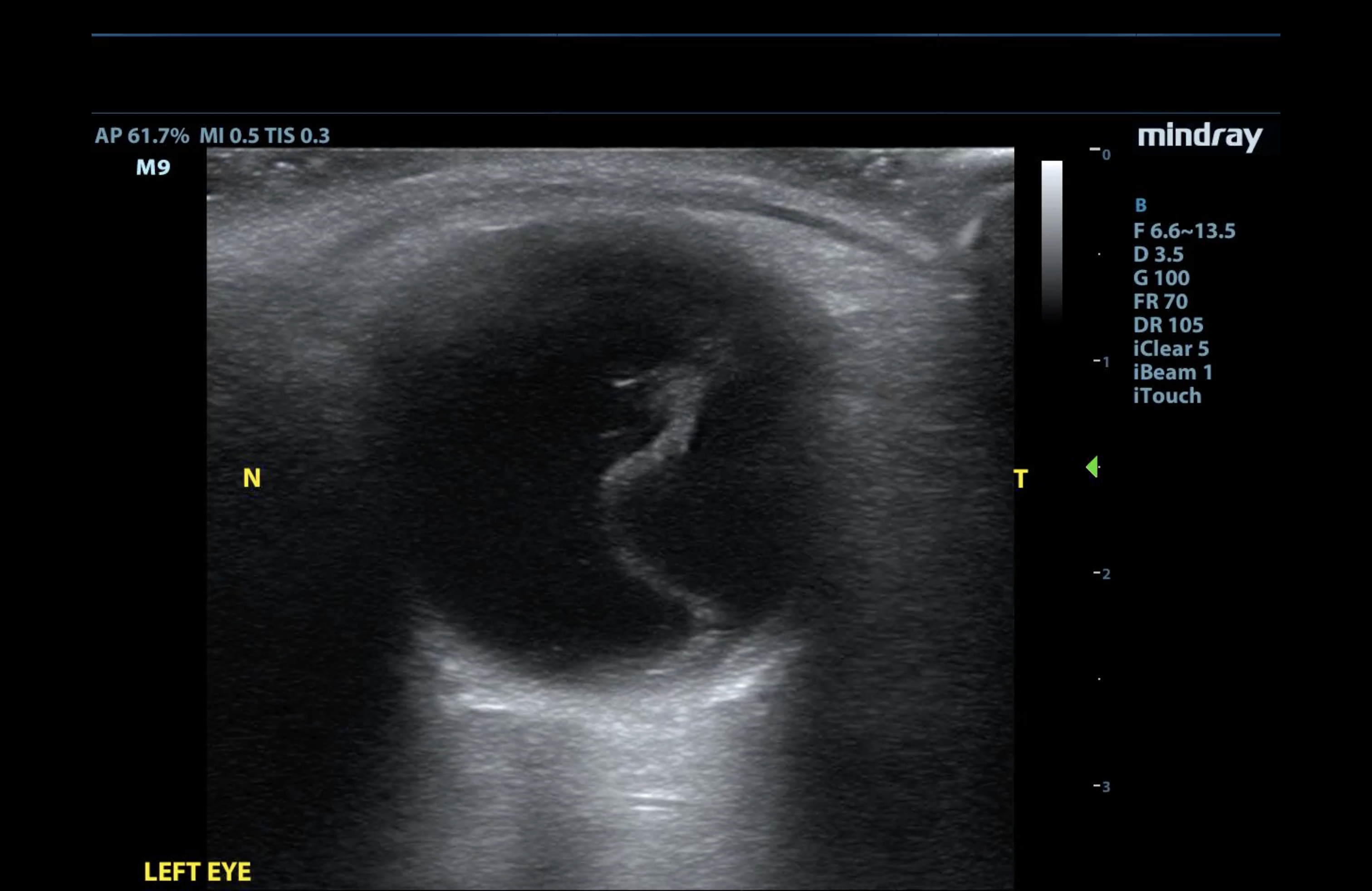}
    \caption{Globe centered in the frame}
    \label{fig:globe_center}
  \end{subfigure}
  \hfill
  \begin{subfigure}[b]{0.45\textwidth}
    \includegraphics[width=\textwidth]{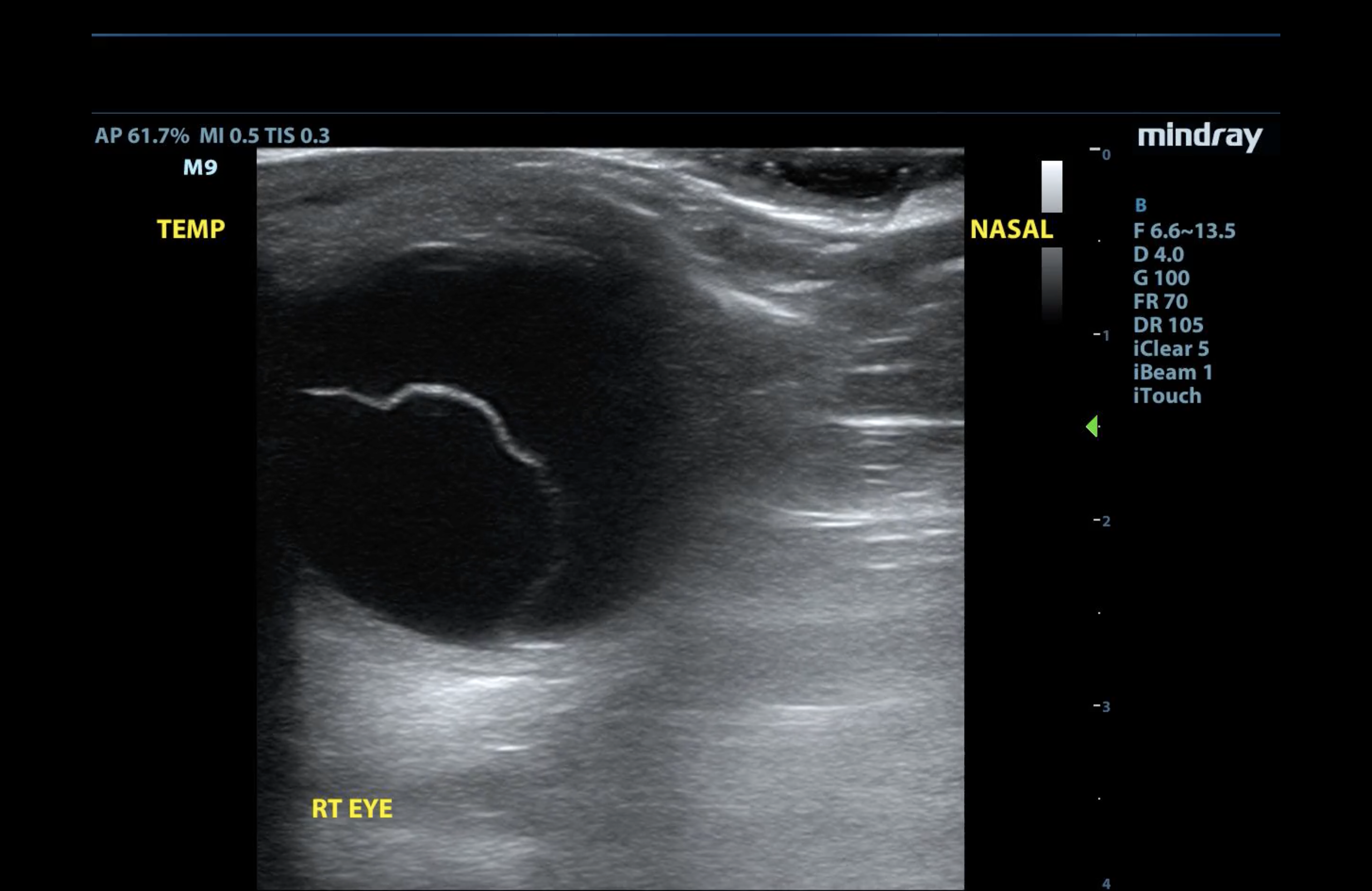}
    \caption{Globe near the periphery of the frame}
    \label{fig:globe_side}
  \end{subfigure}
  \caption{Examples of ocular ultrasound frames with the globe (a) centered and (b) near the image boundary.}
  \label{fig:frame_examples}
\end{figure}

All ultrasound clips underwent mandatory preprocessing prior to release. Although the Qpath export process automatically removes embedded PHI (see Clip Preparation Section), the clips still contain device-generated text overlays (Figure~\ref{fig:eye_anatomy}) that may present residual re-identification risk. These residual texts are also irrelevant---or even detrimental---to automated classification. Since all diagnostic information resides exclusively within the ocular region (our region of interest, ROI; Figure~\ref{fig:eye_anatomy})---including the anterior chamber, lens, vitreous, retina and optic nerve---we initially experimented with a 256 × 256 center crop. Although this was effective for clips where the globe was centrally located, it failed in cases where the globe appeared near the periphery of the image (see Figure~\ref{fig:frame_examples} for examples). Beyond the peripheral globe positioning shown in Figure~\ref{fig:frame_examples}, difficult scenarios include off-axis acquisitions, motion artifacts, variable gain/depth settings, and shadowing/dropout. Consequently, this method was not used in the final pipeline.

\begin{figure}[ht]
  \centering
  \begin{subfigure}[b]{0.45\textwidth}
    \includegraphics[width=\textwidth]{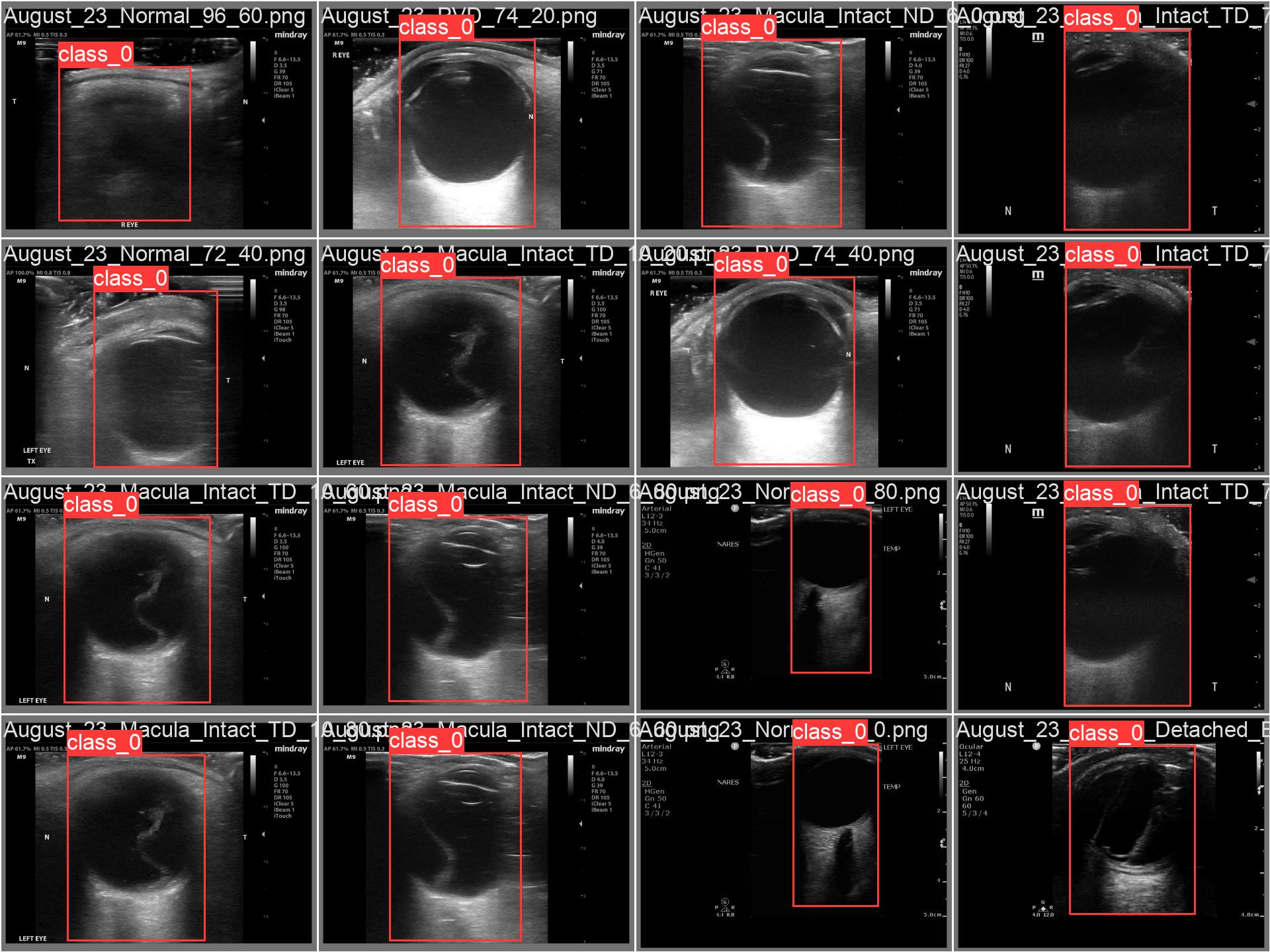}
    \caption{Validation example batch labels}
    \label{fig:yolo_labels}
  \end{subfigure}
  \hfill
  \begin{subfigure}[b]{0.45\textwidth}
    \includegraphics[width=\textwidth]{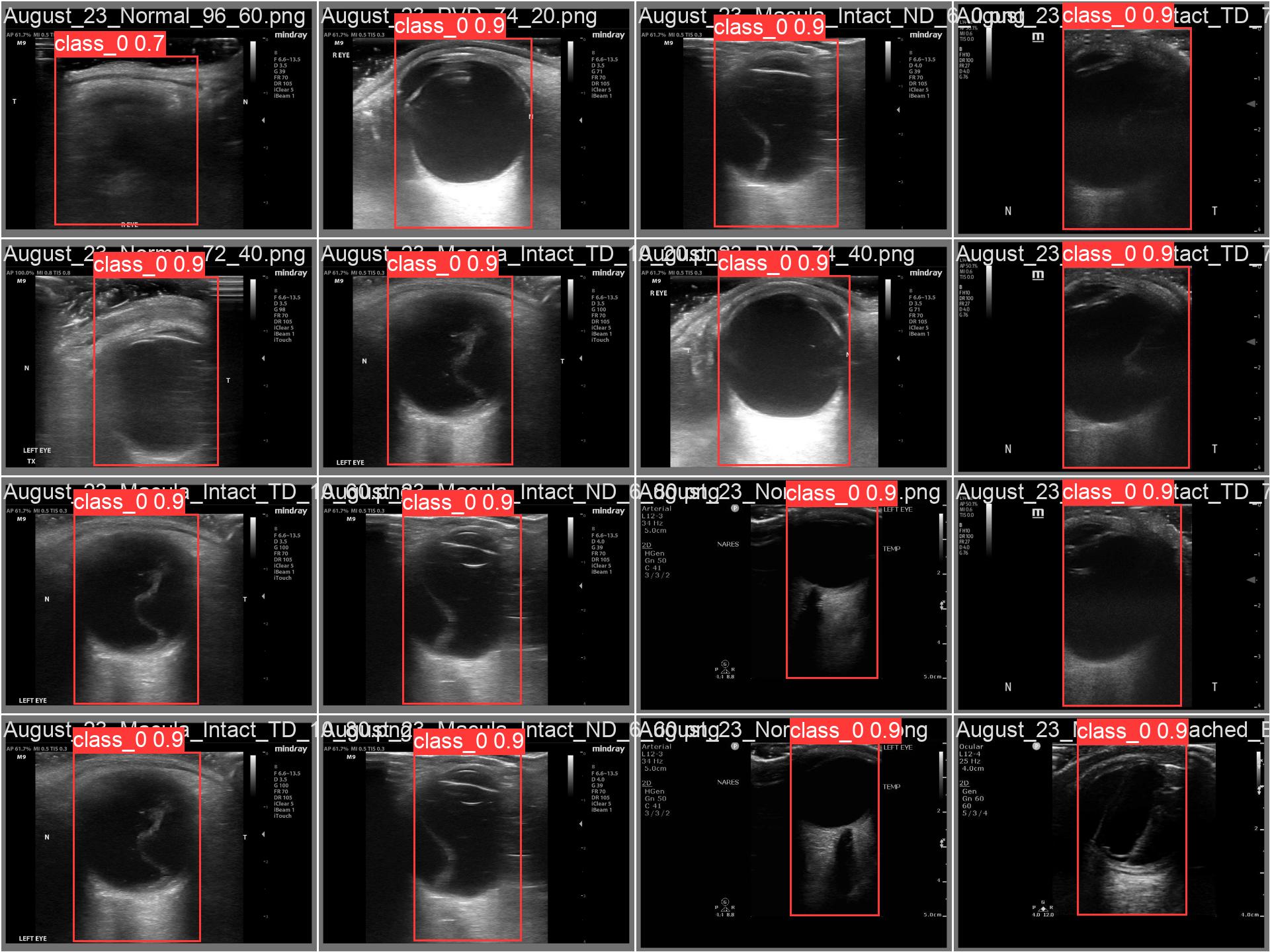}
    \caption{Validation example batch predictions}
    \label{fig:yolo_preds}
  \end{subfigure}
  \caption{Comparison of the example validation batch labels and predictions after YOLOv8 training.}
  \label{fig:yolo_det_results}
\end{figure}

Instead, we adopted a YOLOv8-based preprocessing approach for the entire dataset. We trained the model on 115 manually annotated frames, where bounding boxes were drawn around the ROI (Figure~\ref{fig:eye_anatomy}), and validated it on an additional 51 annotated frames. The model was trained for 100 epochs with hyperparameter tuning. The YOLOv8 model achieved strong performance in the validation set, as shown in Table~\ref{tab:yolo_performance}. Figure~\ref{fig:yolo_det_results} presents a comparison of ground truth labels and YOLOv8 predictions on an example validation batch of 16 different frames from different clips. As shown, the predicted bounding boxes closely match the annotated ground truth, with most predictions achieving confidence scores of 0.9. Even in cases with lower confidence scores (e.g., 0.7), the predicted bounding boxes remain well-aligned with the annotated regions, demonstrating the reliability of the automated ROI detection.

\begin{table}[ht]
\centering
\caption{YOLOv8 ROI detection model performance on the validation set.}
\label{tab:yolo_performance}
\begin{tabular}{cccc}
\toprule
\textbf{Precision} & \textbf{Recall} & \textbf{mAP@50} & \textbf{mAP@50-95} \\
\midrule
0.998 & 1.000 & 0.995 & 0.838 \\
\bottomrule
\end{tabular}
\end{table}


After training, the YOLOv8 model was applied to each video clip to locate the ROI. For each clip, we extracted the bounding boxes from all frames and selected the one with the highest confidence score to define the crop region for the entire clip. This automated preprocessing step is used only to localize the region of interest, allowing classification models to focus exclusively on the relevant anatomical structures.

If the resulting cropped frame dimensions were not divisible by 16, we padded the frame with black pixels. This ensured compatibility with downstream processing and aligned with the architectural constraints of modern Vision Transformer (ViT)-based models, which typically operate on 16×16 patch sizes.

YOLOv8-based preprocessing provides additional safety by cropping the frame to the ocular region of interest, thus removing these peripheral text overlays. The preprocessing does not degrade clip/frame quality, and all diagnostic features reside within the cropped ocular region of interest, ensuring no clinically relevant information is lost.

\subsection*{Data Protection and Compliance}
All ocular ultrasound clips used in this study were thoroughly deidentified, with all Protected Health Information (PHI) removed in accordance with the HIPAA guidelines. The resulting deidentified clip datasets are initially stored in a randomized order on the University of Arizona's HIPAA-compliant Box Health storage platform. The datasets were securely shared with the research team at The Ohio State University (OSU) using OSU SharePoint, a secure, institutionally managed cloud storage system that ensures compliance with data protection standards.

\section*{Data Record}
\begin{figure}[ht]
  \centering
  \begin{subfigure}[b]{0.49\textwidth}
    \includegraphics[width=\textwidth]{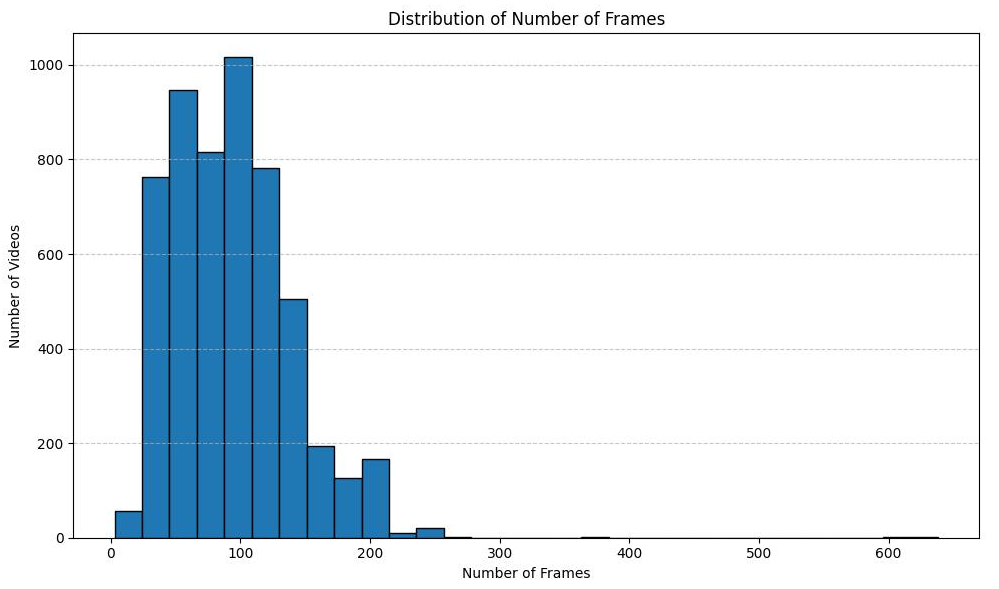}
    \caption{Distribution of number of frames per clip}
    \label{fig:num_frames}
  \end{subfigure}
  \hfill
  \begin{subfigure}[b]{0.49\textwidth}
    \includegraphics[width=\textwidth]{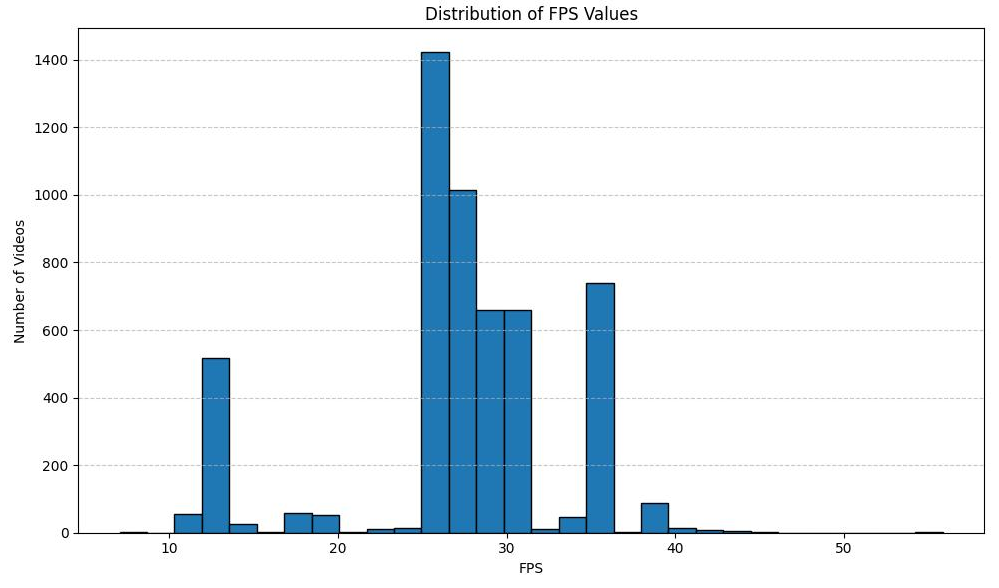}
    \caption{Distribution of FPS values}
    \label{fig:fps_values}
  \end{subfigure}
  \caption{Summary statistics for the ERDES dataset: (a) number of frames per clip and (b) FPS values across all videos.}
  \label{fig:data_stats}
\end{figure}

\begin{figure}[ht]
  \centering
  \includegraphics[height=0.5\textwidth]{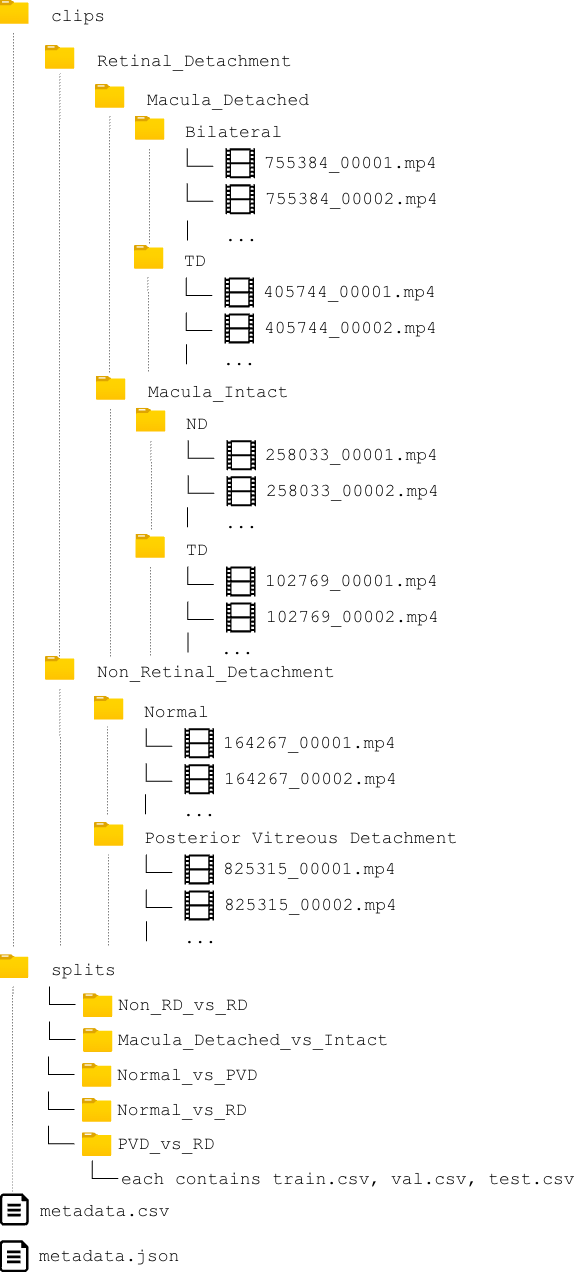}
  \caption{Folder hierarchy of the ERDES dataset.}
  \label{fig:folder_structure}
\end{figure}

The ERDES~\cite{ozkuterdes} dataset is available at Zenodo. It consists of \textbf{5,381} ultrasound clips, totaling approximately \textbf{5 hours and 10 minutes} of video data. Multiple clips may originate from the same patient encounter, as each ocular ultrasound examination typically includes multiple acquisitions from different probe positions and scanning angles. Clip durations range from 0.12 to 25.52 seconds, with a median of 3.1 seconds, an interquartile range (IQR) of 2.1 to 4.2 seconds, and a mean of $3.4 \pm 1.6$ seconds. Figure~\ref{fig:data_stats} summarizes key characteristics of the dataset, including the distribution of video lengths (measured by the number of frames) and frame rates (FPS). A visual overview of the folder hierarchy is provided in Figure~\ref{fig:folder_structure}. This structure enables both high-level binary classification and fine-grained subclass-based analyses. 

The dataset is organized into a hierarchical folder structure that mirrors clinical diagnostic categories. At the top level, video clips are grouped into two primary categories: \texttt{Non\_Retinal\_Detachment} (eyes without retinal detachment) and \texttt{Retinal\_Detachment}. Within the \texttt{Non\_Retinal\_Detachment} category, the clips are further subdivided by \texttt{Posterior\_Vitreous\_Detachment} (PVD) and \texttt{Normal}. Within the \texttt{Retinal\_Detachment} category, the clips are further subdivided by macular status into \texttt{Macula\_Detached} and \texttt{Macula\_Intact}. These subcategories for macular status are then broken down according to the location or extent of detachment into anatomical subclasses: \texttt{Bilateral} and \texttt{TD} (temporal detachment) for \texttt{Macula\_Detached}; and \texttt{ND} (nasal detachment) and \texttt{TD} for \texttt{Macula\_Intact}.

Each video file is named using the format \texttt{\{random\_id\}\_\{counter\}.mp4}, ensuring unique, de-identified clip identifiers with no embedded protected health information (PHI). Labels were recorded at clip level through folder assignment. The total number of clips per diagnostic category and subclass is reported in Table~\ref{tab:class_breakdown}.

\begin{table}[ht]
\centering
\caption{Number of clips in each diagnostic category and anatomical subclass.}
\label{tab:class_breakdown}
\begin{tabular}{l c}
\toprule
\textbf{Class / Subclass} & \textbf{Number of Clips} \\
\midrule
\textbf{Non-Retinal Detachment (Non-RD)} & \textbf{4,879} \\
\quad Normal                      & 4,233 \\
\quad Posterior Vitreous Detachment (PVD)  & 646 \\
\textbf{Retinal Detachment (RD)} & \textbf{502} \\
\quad \textit{Macula Intact}           & 199 \\
\quad \quad Nasal Detachment (ND)     & 88 \\
\quad \quad Temporal Detachment (TD)  & 111 \\
\quad \textit{Macula Detached}         & 303 \\
\quad \quad Temporal Detachment (TD)  & 151 \\
\quad \quad Bilateral Detachment      & 152 \\
\midrule
\textbf{Total}                  & \textbf{5,381} \\
\bottomrule
\end{tabular}
\end{table}

The observed class imbalance in our dataset, with a predominance of Non-RD ocular ultrasound clips compared to cases of RD, is consistent with the clinical prevalence in patients with symptoms related to RD. RD is a relatively rare condition that affects approximately 1 in 10,000 individuals annually, with a lifetime risk near 1 in 300. As a result, most patients who come to the emergency department with visual symptoms do not have RD, and their ultrasound examinations often produce Non-RD findings. This natural prevalence is reflected in our dataset, which includes a higher proportion of Non-RD ocular ultrasound clips. In addition, this distribution aligns with the typical learning trajectory of physician trainees in POCUS. During training, clinicians are disproportionately exposed to Non-RD studies, helping them to build foundational pattern recognition skills before learning to identify less common pathologies such as RD. Similarly, training a machine learning model on a dataset that reflects this clinical distribution is not only realistic, but also pedagogically analogous to human learning, prioritizing exposure to common findings as a basis for developing diagnostic accuracy.

To support reproducibility and ensure proportional class representation despite this imbalance, we provide pre-defined dataset splits for five binary classification tasks: Non-RD vs.\ RD, Macula Detached vs.\ Macula Intact, Normal vs.\ PVD, Normal vs.\ RD, and PVD vs.\ RD under \texttt{splits/} folder. The primary tasks addressed in this article are Non-RD vs.\ RD and Macula Detached vs.\ Macula Intact; the remaining three are included to support broader experimentation with the dataset. The \texttt{train.csv}, \texttt{val.csv}, and \texttt{test.csv} files are included in the dataset release for each task. All splits were generated using stratified sampling with a fixed random seed of 42 and a 72\%/8\%/20\% ratio for training, validation, and test sets, respectively. Detailed class distributions for each split are provided in Tables~\ref{tab:split_rd} and~\ref{tab:split_macula} for the primary tasks, and Table~\ref{tab:splits_supplementary} for the supplementary tasks.

\begin{table}[ht]
\caption{Stratified data splits for primary classification tasks.}
  \centering
  \begin{minipage}{0.45\textwidth}
    \centering
    \begin{tabular}{lccc}
      \toprule
      \textbf{Class} & \textbf{Train} & \textbf{Val} & \textbf{Test} \\
      \midrule
      Non-RD & 3,511 & 391 & 977 \\
      RD     & 362 & 40 & 100 \\
      \midrule
      \textbf{Total} & \textbf{3,873} & \textbf{431} & \textbf{1,077} \\
      \bottomrule
    \end{tabular}
    \subcaption{Non-RD vs.\ RD task}
    \label{tab:split_rd}
  \end{minipage}
  \hfill
  \begin{minipage}{0.45\textwidth}
    \centering
    \begin{tabular}{lccc}
      \toprule
      \textbf{Class} & \textbf{Train} & \textbf{Val} & \textbf{Test} \\
      \midrule
      Macula-Detached  & 217 & 25 & 61 \\
      Macula-Intact & 143 & 16 & 40 \\
      \midrule
      \textbf{Total} & \textbf{360} & \textbf{41} & \textbf{101} \\
      \bottomrule
    \end{tabular}
    \subcaption{Macula-Detached vs.\ Macula-Intact task}
    \label{tab:split_macula}
  \end{minipage}
  \label{tab:splits}
\end{table}

\begin{table}[ht]
\caption{Stratified data splits for supplementary classification tasks.}
  \centering
  \begin{minipage}{0.32\textwidth}
    \centering
    \begin{tabular}{lccc}
      \toprule
      \textbf{Class} & \textbf{Train} & \textbf{Val} & \textbf{Test} \\
      \midrule
      Normal & 3,047 & 339 & 847 \\
      PVD    & 465 & 52 & 129 \\
      \midrule
      \textbf{Total} & \textbf{3,512} & \textbf{391} & \textbf{976} \\
      \bottomrule
    \end{tabular}
    \subcaption{Normal vs.\ PVD task}
    \label{tab:split_normal_pvd}
  \end{minipage}
  \hfill
  \begin{minipage}{0.32\textwidth}
    \centering
    \begin{tabular}{lccc}
      \toprule
      \textbf{Class} & \textbf{Train} & \textbf{Val} & \textbf{Test} \\
      \midrule
      Normal & 3,051 & 337 & 845 \\
      RD     & 362 & 40 & 100 \\
      \midrule
      \textbf{Total} & \textbf{3,413} & \textbf{377} & \textbf{945} \\
      \bottomrule
    \end{tabular}
    \subcaption{Normal vs.\ RD task}
    \label{tab:split_normal_rd}
  \end{minipage}
  \hfill
  \begin{minipage}{0.32\textwidth}
    \centering
    \begin{tabular}{lccc}
      \toprule
      \textbf{Class} & \textbf{Train} & \textbf{Val} & \textbf{Test} \\
      \midrule
      PVD & 465 & 52 & 129 \\
      RD  & 361 & 40 & 101 \\
      \midrule
      \textbf{Total} & \textbf{826} & \textbf{92} & \textbf{230} \\
      \bottomrule
    \end{tabular}
    \subcaption{PVD vs.\ RD task}
    \label{tab:split_pvd_rd}
  \end{minipage}
  \label{tab:splits_supplementary}
\end{table}

Dataset splitting was performed at the clip level rather than the patient level, as patient level identifiers were removed during de-identification. Therefore, multiple clips from the same patient may appear across training, validation, and test sets. Although patient-independent splitting is ideal for strict generalization assessment, each clip represents a distinct acquisition with different probe position, scanning angle, orientation, depth, and dynamic eye motion, resulting in substantial variability in visual appearance across clips. Thus, clips from the same patient are not duplicates and contain different diagnostic information.

The dataset release includes a structured metadata file (\texttt{metadata.csv}) and a dataset summary file (\texttt{metadata.json}). The \texttt{metadata.csv} file contains one row per clip with the following fields: \texttt{clip\_id} (unique identifier), \texttt{file\_path} (relative path to the video file), \texttt{diagnostic\_class} (primary diagnosis: \texttt{rd} or \texttt{non\_rd}), \texttt{subtype} (specific condition: \texttt{normal}, \texttt{pvd}, \texttt{macula\_intact}, or \texttt{macula\_detached}), \texttt{anatomical\_subclass} (location of detachment for RD cases: \texttt{TD}, \texttt{ND}, or \texttt{Bilateral}), and video-level properties including \texttt{fps}, \texttt{frame\_count}, \texttt{width}, \texttt{height}, and \texttt{duration\_seconds}. The \texttt{metadata.json} file provides a machine-readable summary of the dataset that includes label definitions, field descriptions, labeling methodology, and aggregate statistics.

\section*{Technical Validation}
\subsection*{Dataset and Label Validation}

All ultrasound clips in the ERDES dataset were collected between June 2010 and March 2022 at the University of Arizona under protocols approved by the IRB. Each clip was labeled by three ocular ultrasound experts to ensure diagnostic accuracy and reliability. Additional quality control steps were applied to ensure scientific validity and clinical relevance.

\subsection*{Benchmarking and Training Setup}
To benchmark performance, we trained a total of 40 deep learning models based on eight distinct spatiotemporal architectures using the MONAI framework~\cite{Monai}. Each architecture was trained separately for five clinically relevant binary classification tasks: (i) eyes without retinal detachment (Non-RD) vs.\ retinal detachment (RD), (ii) Normal vs.\ RD, (iii) PVD vs.\ RD, (iv) Macula Detached vs.\ Macula Intact, and (v) Normal vs.\ PVD. All models were initialized from scratch, without external pretraining, to evaluate performance solely on ERDES data.

Prior to model training, each cropped clip underwent additional preprocessing: conversion to grayscale, symmetric zero-padding to square, spatiotemporal resizing via trilinear interpolation, and intensity normalization. Table~\ref{tab:preprocessing} summarizes the preprocessing steps from raw video properties to the format used as input to model training, validation, and testing.

\begin{table}[htbp]
\centering
\caption{Video Pre-processing Pipeline and Post-processed Statistics}
\label{tab:preprocessing}
\begin{tabular}{ll}
\toprule
\textbf{Property} & \textbf{Value} \\
\midrule
\multicolumn{2}{l}{\textit{Raw Video Properties}} \\
Spatial resolution (W$\times$H) & 272 $\times$ 272 to 1088 $\times$ 1088 \\
Frame count & 3--638 (mean: 92) \\
Frame rate (FPS) & 7.0--55.9 (mean: 26.9) \\
Duration (seconds) & 0.12--25.52 (mean: 3.44) \\
Mean pixel intensity & 27.0 $\pm$ 42.6\\
\midrule
\multicolumn{2}{l}{\textit{Pre-processing Pipeline}} \\
Color conversion & RGB $\rightarrow$ grayscale (channel mean) \\
Padding strategy & Symmetric zero-padding to square \\
Spatiotemporal resize & Trilinear interpolation to 96$\times$128$\times$128 (D$\times$H$\times$W) \\
\midrule
\multicolumn{2}{l}{\textit{Post-processed Properties}} \\
Output dimensions (D$\times$H$\times$W) & 96 $\times$ 128 $\times$ 128 \\
Channels & 1 (grayscale) \\
Normalization (pixel range) & [0, 255] $\rightarrow$ [0, 1] (divide by 255) \\
Mean pixel intensity (normalized) & 0.106 $\pm$ 0.167 \\
\bottomrule
\end{tabular}
\end{table}

The selected architectures span both convolutional and transformer-based paradigms and include 3D ResNet~\cite{3dresnet}, 3D U-Net~\cite{cicekunet3d}, V-Net~\cite{milletari2016v}, UNet++~\cite{zhou2018unet++}, SENet154~\cite{hu2018senet}, Swin-UNETR~\cite{hatamizadeh2021swin}, UNETR~\cite{hatamizadeh2022unetr}, and a ViT~\cite{dosovitskiy2020image}-based classifier. The eight architectures were trained on the five binary tasks on a compute node equipped with three NVIDIA A6000 GPUs (48GB each). We used the AdamW optimizer with an initial learning rate of $1.5 \times 10^{-5}$. Each model was trained for 50 epochs. Binary cross-entropy with logit loss was used as the loss function during training and validation. Batch sizes varied across architectures depending on GPU memory requirements and the model FLOP count, typically ranging from a total of 12 to 96 across the three GPUs, with larger models using smaller batch sizes to fit in memory.

No class weighting or resampling was applied as we aimed to evaluate model performance under realistic class distributions. Instead, for each binary classification task, we used the pre-defined stratified splits (described in the Data Records section) to ensure proportional class representation across training, validation, and test sets. The Non-RD vs.\ RD split was constructed using all clips from the \texttt{Non\_Retinal\_Detachment/} and \texttt{Retinal\_Detachment/} directories, including all subfolders. For the Macula Detached vs.\ Macula Intact task, all clips within the \texttt{Retinal\_Detachment/} directory were included and grouped into \texttt{Macula\_Detached} and \texttt{Macula\_Intact} according to the macular labels indicated by their parent subfolder. The remaining three tasks (Normal vs.\ PVD, Normal vs.\ RD, and PVD vs.\ RD) were each constructed by selecting clips from the corresponding subdirectories. In all five tasks, anatomical subclasses such as \texttt{TD}, \texttt{ND}, and \texttt{Bilateral} were not treated as separate classes but were always grouped under the corresponding binary label. Each sample was labeled with a binary class as shown in Table~\ref{tab:class_labels}.

\begin{table}[ht]
  \centering
  \caption{Binary class assignments for each classification task.}
  \label{tab:class_labels}
  \begin{tabular}{lcc}
    \toprule
    \textbf{Task} & \textbf{Class 0} & \textbf{Class 1} \\
    \midrule
    Non-RD vs.\ RD              & Non-RD          & RD \\
    Macula Detached vs.\ Intact & Macula Detached & Macula Intact \\
    Normal vs.\ PVD             & Normal           & PVD \\
    Normal vs.\ RD              & Normal           & RD \\
    PVD vs.\ RD                 & PVD              & RD \\
    \bottomrule
  \end{tabular}
\end{table}
 
Performance was evaluated using five standard classification metrics, namely \textit{Accuracy}, \textit{Precision}, \textit{Sensitivity (Recall)}, \textit{Specificity}, and the \textit{F1-Score}. These metrics provide complementary insights into the performance of model classification, especially in the context of imbalanced datasets.

Accuracy measures the overall correctness of the model by evaluating the proportion of true results (both true positives and true negatives) among all predictions.

\begin{equation}
\mathrm{Accuracy} = \frac{TP + TN}{TP + TN + FP + FN}
\label{eq:accuracy}
\end{equation}

Precision measures the fraction of positive predictions that correspond to true positive cases, indicating the accuracy of the positive classifications of the model.

\begin{equation}
\mathrm{Precision} = \frac{TP}{TP + FP}
\label{eq:precision}
\end{equation}

Sensitivity (also known as Recall) assesses the ability of the model to identify all actual positive cases. It is especially important in scenarios where false negatives are costly.

\begin{equation}
\mathrm{Sensitivity} = \frac{TP}{TP + FN}
\label{eq:recall}
\end{equation}

Specificity complements sensitivity by measuring the proportion of actual negative cases that are correctly identified.

\begin{equation}
\mathrm{Specificity} = \frac{TN}{TN + FP}
\label{eq:specificity}
\end{equation}

The F1-score provides the harmonic mean of precision and recall and balances the two metrics, especially useful when the class distribution is imbalanced.

\begin{equation}
   \mathrm{F1\text{-}Score} = \frac{2 \cdot \mathrm{Precision} \cdot \mathrm{Recall}}{\mathrm{Precision} + \mathrm{Recall}}
\label{eq:f1} 
\end{equation}

Table~\ref{tab:rd_results}, Table~\ref{tab:macula_results} and Table~\ref{tab:normal_pvd_pooling_comparison_4way} present the model performance in task-specific test splits for the five binary classification tasks. The primary evaluations focus on Non-RD vs.\ RD and Macula Detached vs.\ Macula Intact, which reflect the intended clinical use cases. The remaining three tasks (Normal vs.\ RD, PVD vs.\ RD, and Normal vs.\ PVD) provide a broader diagnostic context and illustrate model behavior on clinically related but more ambiguous distinctions, particularly given that PVD can mimic RD on ultrasound.

\begin{table}[ht]
\caption{Performance of models on three retinal detachment (RD) classification tasks: Normal vs.\ RD, PVD vs.\ RD, and Non-RD vs.\ RD. Metrics include sensitivity (Sens), specificity (Spec), precision (Prec), F1-score (F1), and accuracy (Acc). Models are sorted in ascending order based on sensitivity for the Normal vs.\ RD task. The best-performing model based on sensitivity for each task is shown in \textcolor{blue}{\textbf{bold blue}}, and the second-best is highlighted in \textcolor{orange}{orange}.}
  \centering
  \begin{adjustbox}{max width=\textwidth}
  \begin{tabular}{lccccc|ccccc|ccccc}
    \toprule
    \multirow{2}{*}{\textbf{Backbone}} &
      \multicolumn{5}{c|}{\textbf{Normal vs.\ RD}} &
      \multicolumn{5}{c|}{\textbf{PVD vs.\ RD}} &
      \multicolumn{5}{c}{\textbf{Non-RD vs.\ RD}} \\
    \cmidrule(r){2-6} \cmidrule(lr){7-11} \cmidrule(l){12-16}
          & \textbf{Sens} & \textbf{Spec} & \textbf{Prec} & \textbf{F1} & \textbf{Acc}
          & \textbf{Sens} & \textbf{Spec} & \textbf{Prec} & \textbf{F1} & \textbf{Acc}
          & \textbf{Sens} & \textbf{Spec} & \textbf{Prec} & \textbf{F1} & \textbf{Acc} \\
    \midrule
    ViT~\cite{dosovitskiy2020image}        & 0.643 & 0.971 & 0.730 & 0.684 & 0.937 & 0.676 & 0.790 & 0.718 & 0.696 & 0.740 & 0.689 & 0.969 & 0.696 & 0.693 & 0.943 \\
    UNETR~\cite{hatamizadeh2022unetr}      & 0.702 & 0.975 & 0.771 & 0.735 & 0.946 & 0.617 & 0.883 & 0.807 & 0.699 & 0.766 & 0.569 & 0.972 & 0.678 & 0.619 & 0.935 \\
    SENet154~\cite{hu2018senet}   & 0.732 & 0.995 & 0.948 & 0.826 & 0.967 & 0.705 & 0.961 & 0.935 & 0.804 & 0.848 & 0.689 & 0.992 & 0.907 & 0.784 & 0.964 \\
    Swin-UNETR~\cite{hatamizadeh2021swin} & 0.742 & 0.971 & 0.757 & 0.750 & 0.947 & 0.774 & 0.744 & 0.705 & 0.738 & 0.757 & 0.680 & 0.980 & 0.781 & 0.727 & 0.952 \\
    V-Net~\cite{milletari2016v}      & 0.831 & 0.997 & 0.976 & 0.898 & 0.980 & 0.794 & 0.860 & 0.818 & 0.805 & 0.831 & 0.829 & 0.994 & 0.943 & 0.882 & 0.979 \\
    3D ResNet~\cite{3dresnet}  & 0.900 & 0.987 & 0.892 & 0.896 & 0.977 & 0.882 & 0.875 & 0.849 & 0.865 & 0.878 &
      \textcolor{blue}{\textbf{0.939}} & \textcolor{blue}{\textbf{0.978}} & \textcolor{blue}{\textbf{0.817}} & \textcolor{blue}{\textbf{0.874}} & \textcolor{blue}{\textbf{0.974}} \\
    UNet++~\cite{zhou2018unet++}     & \textcolor{orange}{0.910} & \textcolor{orange}{0.994} & \textcolor{orange}{0.948} & \textcolor{orange}{0.929} & \textcolor{orange}{0.985} &
      \textcolor{blue}{\textbf{0.921}} & \textcolor{blue}{\textbf{0.860}} & \textcolor{blue}{\textbf{0.839}} & \textcolor{blue}{\textbf{0.878}} & \textcolor{blue}{\textbf{0.887}} &
      0.879 & 0.986 & 0.871 & 0.875 & 0.976 \\
    3D U-Net~\cite{cicekunet3d}   & \textcolor{blue}{\textbf{0.950}} & \textcolor{blue}{\textbf{0.996}} & \textcolor{blue}{\textbf{0.969}} & \textcolor{blue}{\textbf{0.959}} & \textcolor{blue}{\textbf{0.991}} &
      \textcolor{orange}{0.901} & \textcolor{orange}{0.852} & \textcolor{orange}{0.828} & \textcolor{orange}{0.863} & \textcolor{orange}{0.874} &
      \textcolor{orange}{0.920} & \textcolor{orange}{0.988} & \textcolor{orange}{0.893} & \textcolor{orange}{0.906} & \textcolor{orange}{0.982} \\
    \bottomrule
  \end{tabular}
  \end{adjustbox}
  \label{tab:rd_results}
\end{table}

\begin{table}[ht]
\caption{Performance of models on macular status classification (macula-detached vs.\ -intact). Metrics include sensitivity (Sens), specificity (Spec), precision (Prec), F1-score (F1), and accuracy (Acc). The best-performing model based on sensitivity is shown in \textcolor{blue}{\textbf{bold blue}}, and the second-best is highlighted in \textcolor{orange}{orange}. Models follow the same ordering as in Table~\ref{tab:rd_results} to facilitate cross-task comparison.}
  \centering
  \begin{adjustbox}{max width=0.75\textwidth}
  \begin{tabular}{lccccc}
    \toprule
    \multirow{2}{*}{\textbf{Backbone}} &
      \multicolumn{5}{c}{\textbf{Macula-Detached vs.\ Intact}} \\
    \cmidrule(l){2-6}
          & \textbf{Sens} & \textbf{Spec} & \textbf{Prec} & \textbf{F1} & \textbf{Acc} \\
    \midrule
    ViT        & 0.699 & 0.741 & 0.636 & 0.666 & 0.725 \\
    UNETR      & 0.774 & 0.709 & 0.632 & 0.696 & 0.735 \\
    SENet154   & 0.625 & 0.903 & 0.806 & 0.704 & 0.794 \\
    Swin-UNETR & 0.699 & 0.806 & 0.699 & 0.699 & 0.764 \\
    V-Net      & \textcolor{orange}{0.824} & \textcolor{orange}{0.806} & \textcolor{orange}{0.733} & \textcolor{orange}{0.776} & \textcolor{orange}{0.813} \\
    3D ResNet  & 0.750 & 0.935 & 0.882 & 0.810 & 0.862 \\
    UNet++     & 0.699 & 0.967 & 0.933 & 0.800 & 0.862 \\
    3D U-Net   & \textcolor{blue}{\textbf{0.899}} & \textcolor{blue}{\textbf{0.870}} & \textcolor{blue}{\textbf{0.818}} & \textcolor{blue}{\textbf{0.857}} & \textcolor{blue}{\textbf{0.882}} \\
    \bottomrule
  \end{tabular}
  \end{adjustbox}
  \label{tab:macula_results}
\end{table}

Across all tasks, models use 3D global average pooling over the full temporal extent of each clip as the default pooling strategy. However, for the Normal vs.\ PVD task, this approach yielded lower performance compared to the other tasks. We attribute this to the nature of PVD, which is more mobile with eye movements and less echogenic than RD, making it visible in fewer frames of a given clip. Using all frames equally may therefore dilute the diagnostically relevant signal. To validate that this lower performance reflects genuine temporal characteristics of PVD rather than label noise, we explored a selective pooling strategy that operates on 2D spatial features rather than the full 3D volume.

Each backbone produces a different number of temporal segments $D$ from the input clip of 96 frames, depending on its architecture (e.g. $D=6$ for 3D U-Net, $D=12$ for UNETR). Rather than pooling across all $D$ segments equally, the selective strategy scores each segment and retains only the top $k^{\ast} = \lceil D \cdot r \rceil$ segments, where $r$ is a selection ratio. We evaluated three selection ratios: $r \in \{0.7, 0.5, 0.3\}$, corresponding to retaining approximately 70\%, 50\%, and 30\% of the temporal segments, respectively. The ceiling function ensures that at least one segment is always selected. When $\lceil D \cdot r \rceil = D$, all segments are retained and the selective strategy reduces to the standard global pooling; these cases are marked with dashes in the table. For backbones with very few segments (e.g., ViT with $D=1$), selective pooling is not applicable at any ratio, as there is only a single segment to select from.

Table~\ref{tab:normal_pvd_pooling_comparison_4way} compares these selective strategies with global average pooling for the Normal vs.\ PVD task. The goal is to determine whether focusing on a subset of temporally informative segments can improve the detection of PVD, which appears in fewer frames and with lower echogenicity compared to RD. The results show that selective pooling can improve performance over global pooling for most backbones; however, the optimal selection ratio varies between architectures. For example, 3D ResNet benefits the most from $r=0.7$, while UNet++ and 3D U-Net achieve their best sensitivity at $r=0.3$. This suggests that the ideal ratio depends on the temporal resolution and feature characteristics of each backbone and that no single selection ratio is universally optimal.

\begin{table}[ht]
\caption{Comparison of temporal pooling strategies for Normal vs.\ PVD classification. All input clips consist of 96 frames, which are grouped into $D$ temporal segments prior to classification. Selective pooling selects the top $k^\ast = \lceil D \cdot r \rceil$ segments, where $r$ is the selection ratio. For each backbone, the best-performing pooling strategy (by sensitivity) is shown in \textbf{bold}. Within each pooling strategy, the best-performing backbone is highlighted in \textcolor{blue}{blue} and the second-best in \textcolor{orange}{orange}, both determined by sensitivity. Dashes (--) indicate configurations where $\lceil D \cdot r \rceil = D$, meaning all segments are retained and no selection occurs.}
\centering
\begin{adjustbox}{max width=\textwidth}
\begin{tabular}{lc|ccccc|ccccc|ccccc|ccccc}
\toprule
\multirow{2}{*}{\textbf{Backbone}} &
\multirow{2}{*}{\textbf{$D$}} &
\multicolumn{5}{c|}{\textbf{Avg Global Pooling}} &
\multicolumn{5}{c|}{\textbf{Selective ($k^\ast{=}\lceil D*0.7 \rceil$)}} &
\multicolumn{5}{c|}{\textbf{Selective ($k^\ast{=}\lceil D*0.5 \rceil$)}} &
\multicolumn{5}{c}{\textbf{Selective ($k^\ast{=}\lceil D*0.3 \rceil$)}} \\
\cmidrule(lr){3-7} \cmidrule(lr){8-12} \cmidrule(lr){13-17} \cmidrule(lr){18-22}
 & & \textbf{Sens} & \textbf{Spec} & \textbf{Prec} & \textbf{F1} & \textbf{Acc}
 & \textbf{Sens} & \textbf{Spec} & \textbf{Prec} & \textbf{F1} & \textbf{Acc}
 & \textbf{Sens} & \textbf{Spec} & \textbf{Prec} & \textbf{F1} & \textbf{Acc}
 & \textbf{Sens} & \textbf{Spec} & \textbf{Prec} & \textbf{F1} & \textbf{Acc} \\
\midrule
ViT        & 1
           & \textbf{0.453} & \textbf{0.964} & \textbf{0.662} & \textbf{0.538} & \textbf{0.896}
           & -- & -- & -- & -- & --
           & -- & -- & -- & -- & --
           & -- & -- & -- & -- & -- \\
UNETR      & 12
           & 0.453 & 0.974 & 0.728 & 0.559 & 0.904
           & 0.484 & 0.965 & 0.684 & 0.567 & 0.901
           & 0.553 & 0.950 & 0.631 & 0.590 & 0.897
           & \textbf{0.561} & \textbf{0.938} & \textbf{0.583} & \textbf{0.572} & \textbf{0.888} \\
SENet154   & 3
           & 0.569 & 0.968 & 0.732 & 0.640 & 0.915
           & -- & -- & -- & -- & --
           & 0.530 & 0.969 & 0.726 & 0.613 & 0.911
           & \textbf{0.584} & \textbf{0.965} & \textbf{0.723} & \textbf{0.646} & \textbf{0.915} \\
Swin-UNETR & 3
           & 0.415 & 0.975 & 0.720 & 0.526 & 0.900
           & -- & -- & -- & -- & --
           & \textbf{0.538} & \textbf{0.961} & \textbf{0.679} & \textbf{0.600} & \textbf{0.904}
           & 0.530 & 0.963 & 0.689 & 0.600 & 0.905 \\
V-Net      & 6
           & \textbf{0.500} & \textbf{0.969} & \textbf{0.714} & \textbf{0.588} & \textbf{0.906}
           & 0.453 & 0.961 & 0.641 & 0.531 & 0.893
           & 0.492 & 0.959 & 0.653 & 0.561 & 0.897
           & 0.446 & 0.966 & 0.674 & 0.537 & 0.897 \\
3D ResNet  & 6
           & \textcolor{orange}{0.638} & \textcolor{orange}{0.982} & \textcolor{orange}{0.846} & \textcolor{orange}{0.728} & \textcolor{orange}{0.936}
           & \textcolor{orange}{\textbf{0.723}} & \textcolor{orange}{\textbf{0.982}} & \textcolor{orange}{\textbf{0.862}} & \textcolor{orange}{\textbf{0.786}} & \textcolor{orange}{\textbf{0.947}}
           & 0.607 & 0.983 & 0.849 & 0.708 & 0.933
           & 0.476 & 0.983 & 0.815 & 0.601 & 0.916 \\
UNet++     & 6
           & 0.492 & 0.981 & 0.800 & 0.609 & 0.916
           & 0.676 & 0.974 & 0.800 & 0.733 & 0.934
           & \textcolor{orange}{0.630} & \textcolor{orange}{0.982} & \textcolor{orange}{0.845} & \textcolor{orange}{0.722} & \textcolor{orange}{0.935}
           & \textcolor{orange}{\textbf{0.699}} & \textcolor{orange}{\textbf{0.983}} & \textcolor{orange}{\textbf{0.866}} & \textcolor{orange}{\textbf{0.774}} & \textcolor{orange}{\textbf{0.945}} \\
3D U-Net   & 6
           & \textcolor{blue}{0.784} & \textcolor{blue}{0.985} & \textcolor{blue}{0.894} & \textcolor{blue}{0.836} & \textcolor{blue}{0.959}
           & \textcolor{blue}{0.800} & \textcolor{blue}{0.983} & \textcolor{blue}{0.881} & \textcolor{blue}{0.838} & \textcolor{blue}{0.959}
           & \textcolor{blue}{0.800} & \textcolor{blue}{0.976} & \textcolor{blue}{0.838} & \textcolor{blue}{0.818} & \textcolor{blue}{0.952}
           & \textcolor{blue}{\textbf{0.807}} & \textcolor{blue}{\textbf{0.977}} & \textcolor{blue}{\textbf{0.846}} & \textcolor{blue}{\textbf{0.826}} & \textcolor{blue}{\textbf{0.955}} \\
\bottomrule
\end{tabular}
\end{adjustbox}
\label{tab:normal_pvd_pooling_comparison_4way}
\end{table}

The ability of multiple architectures to learn meaningful patterns across all five tasks suggests that the ERDES dataset provides sufficiently discriminative features for a range of ocular ultrasound classification tasks. Future work may further break down performance by anatomical subclass (e.g., \texttt{TD}, \texttt{ND}, \texttt{Bilateral}) for finer-grained evaluation, explore different training strategies and architectures for improved performance, and investigate learnable temporal segment selection mechanisms that can adapt the selection ratio per backbone automatically.

\subsection*{Two-Stage Diagnostic Pipeline}
\begin{figure}[ht]
  \centering
  \includegraphics[width=1 \textwidth]{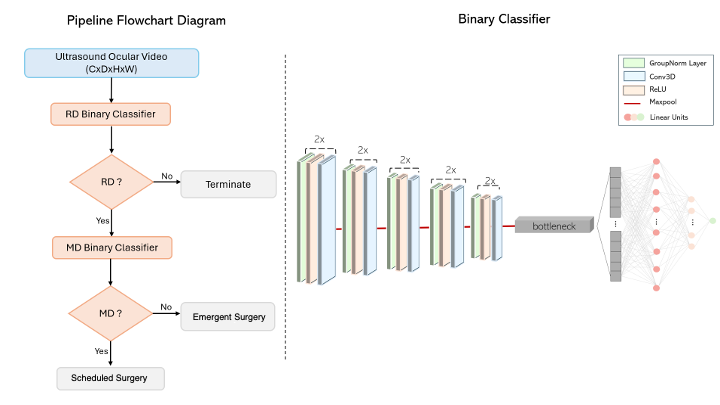}
  \caption{\textbf{Our proposed deep learning-based diagnostic workflow for retinal detachment using ocular ultrasound videos.} The pipeline begins with an input ocular ultrasound video of dimensions C×D×H×W, which is fed into a binary classification model to determine retinal attachment status. If the retina is classified as attached, the process terminates. If detached (RD), the video proceeds to a second classifier that assesses the macular status, categorizing the case as either Macula-Intact or Macula-Detached. Macula-intact patients are referred to the ophthalmology for emergent surgery, while macula-detached cases are associated with reduced visual prognosis and are referred for less urgent intervention. The right side of the figure illustrates the deep learning architecture used for classification.}
  \label{fig:pipeline}
\end{figure}

In a cascaded diagnostic approach for patients presenting with visual symptoms, the sensitivity and specificity of ocular ultrasound are first used to determine the presence of retinal detachment (RD). Among those diagnosed with RD, these test characteristics remain relevant for assessing whether the macula is intact or detached. Our pipeline (Figure~\ref{fig:pipeline}) mimics the clinical decision-making process typically followed in the evaluation of RD. First, ophthalmologists determine whether there is retinal detachment (RD). If RD is detected, the next step is to assess the macular status, specifically whether the macula is intact or detached. This sequential diagnosis was operationalized through a cascaded two-stage classification system:

\begin{enumerate}
  \item \textbf{Stage 1---Retinal Detachment Detection.}  
        The diagnostic pipeline begins with the detection of the presence or absence of retinal detachment (RD), reflecting the first step in clinical workflows in the real world. Each ultrasound video clip is processed as a spatiotemporal volume by our trained model to distinguish RD from non-RD anatomy. This model outputs a probability score $\mathcal{P}(\mathrm{RD})$, and a diagnosis of RD is assigned if $\mathcal{P}(\mathrm{RD}) \ge 0.5$.

  \item \textbf{Stage 2---Macular Status Classification.}  
        If retinal detachment is detected, the pipeline proceeds to a second stage to assess macular involvement, which is one of the most urgent and vision-critical distinctions. A second model classifies the same spatiotemporal volume to determine whether the macula is still intact (\texttt{Macula\_Intact}) or has detached (\texttt{Macula\_Detached}). This decision is again made using a sigmoid threshold: $\mathcal{P}(\mathrm{macula\_detached}) \ge 0.5$ results in a \texttt{Macula\_Detached} diagnosis.

\end{enumerate}

Additionally, when Stage 1 classifies a clip as Non-RD, the pipeline can optionally be extended to distinguish Normal eyes from PVD, in addition to macular status classification. Although this secondary classification is not the primary focus of the current work, the ERDES dataset supports this task, and the benchmarking results in Table~\ref{tab:normal_pvd_pooling_comparison_4way} demonstrate its feasibility.

Based on our benchmarking results (see Table~\ref{tab:rd_results} and Table~\ref{tab:macula_results}), we selected the model trained on 3D ResNet architecture for Stage 1 (Non-RD vs.\ RD task) and the model trained on 3D U-Net architecture for Stage 2 (Macula-Detached vs.\ -Intact task) due to their task-related superior sensitivity performance. It is important to note that the two-stage pipeline is inherently sequential, meaning Stage 2 performance is conditional on Stage 1 correctness. Stage 2 classifies macular status only for clips predicted as RD-positive by Stage 1; therefore, the Stage 2 metrics reported in Table~\ref{tab:macula_results} represent the performance given the correct RD-positive classification. End-to-end pipeline performance is influenced by two sources of error propagation: (i) Stage 1 false negatives (missed RDs), which completely prevent downstream macular classification, and (ii) Stage 1 false positives (Non-RD clips incorrectly classified as RD), which propagate clinically irrelevant cases into Stage 2. This dependency reflects the intended clinical workflow, where the assessment of macular status is only meaningful after the RD has been confirmed. To quantify end-to-end performance, we computed the combined sensitivity by multiplying Stage 1 sensitivity by Stage 2 sensitivity (macula-intact) or Stage 2 specificity (macula-detached): the pipeline achieves an end-to-end sensitivity of 0.844 (0.939 $\times$ 0.899) for detecting macula-intact RD cases and 0.816 (0.939 $\times$ 0.870) for detecting macula-detached RD cases. Given the high Stage 1 sensitivity (0.939) and specificity (0.978), error propagation remains limited, although these combined metrics reflect the true clinical detection capability of the full pipeline.

\section*{Data availability}
The ERDES dataset~\cite{ozkuterdes}, including all ultrasound video clips, pre-defined splits, and metadata files, is publicly available on Zenodo at \url{https://doi.org/10.5281/zenodo.18644370}.

\section*{Code availability}
All scripts for data preprocessing, experiments, model implementation, and configuration files are available in our official GitHub repository at \url{https://github.com/OSUPCVLab/ERDES}. This repository contains the complete codebase for reproducing our experiments and for further model development.

\section*{Author Contributions}
Yasemin Ozkut was the primary technical contributor responsible for the execution of the experiments, data analysis, the development of the selective temporal pooling strategy, dataset preparation and deposition, and the preparation of the manuscript, including the draft of the initial version. Pouyan Navard contributed to the development of the deep learning model, the design of the experimental framework, data analysis, the implementation of the training pipeline, and the preparation of the manuscript. Srikar Adhikari contributed to the conception of the study and assisted in the collection, curation, classification, model development, and manuscript preparation. Alper Yilmaz served as the technical lead, overseeing the project, helping with experimental design, data engineering, manuscript preparation, and mentoring Yasemin Ozkut and Pouyan Navard. Elaine Situ-LaCasse, Josie Acuña and Adrienne A. Yarnish were responsible for the extraction, curation, classification, and labeling of ultrasound data.

\section*{Competing interests}

The authors declare that they have no competing interests. There are no financial, personal or professional conflicts that could have influenced the design, execution, interpretation, or publication of this research. The authors affirm that the integrity of the research was not compromised by external interests or relationships and that all contributions were made with the aim of advancing scientific knowledge.

\section*{Funding}

This work was funded by the Congressionally Directed Medical Research Program (CDMRP) mechanism of the Department of Defense, US Army Medical Research and Development Command, under Grant Number W81XWH-22-1-0803 (Grant Log No. VR210132).

\end{document}